\documentclass[a4paper,11pt]{article}
\usepackage{atsirk}
\usepackage[cp1251]{inputenc}%for Windows cyrillic encoding
\usepackage[russian]{babel}
\usepackage{graphicx}
\usepackage{amsmath}
\usepackage{amssymb}
\usepackage{amstext}
\usepackage{hyperref}

\hypersetup{
    colorlinks=true,
    linkcolor=blue,
    filecolor=magenta,      
    urlcolor=cyan,
    pdftitle={Overleaf Example},
    pdfpagemode=FullScreen,
    }

\begin{document}
\makeatletter
\renewcommand{\@oddhead}{}
\renewcommand{\@evenhead}{}
\renewcommand{\@evenfoot}{\hbox to \textwidth {Astron. Tsirkulyar\textnumero~1658\hfil\thepage\hfil July 2023}}
\renewcommand{\@oddfoot}{\hbox to \textwidth {Astron. Tsirkulyar \textnumero~1658\hfil\thepage\hfil July 2023}}
\renewcommand{\figurename}{Figure}% for text in English
\renewcommand{\tablename}{Table}% for text in English
\makeatother
{ISSN 0236-2457}  \hfill {DOI:10.24412/0236-2457-2023-1658-1-4}

{\large\bf
\centerline{ASTRONOMICHESKII TSIRKULYAR}}
\medskip
\hrule
\medskip
\large
\centerline{Published by the Eurasian Astronomical Society}
\centerline{and Sternberg Astronomical Institute}
\medskip
\hrule
\medskip
\centerline{\textnumero\ 1658, July 2023}
\medskip
\hrule
\bigskip
\centerline{\textbf{Estimates of the Height and Date of the 25th Cycle of Solar Activity}}
%\centerline{\textbf{Capitalize All the Words,}}
%\centerline{\textbf{except for Prepositions and Articles}}
\medskip
\centerline{\textbf{V.N.~Obridko$^1$, D.D.~Sokoloff$^2$, and
M.M.~Katsova$^3$}}
\smallskip
\centerline{\textit{$^1$IZMIRAN, Troitsk, Moscow, Russia}}
\centerline{\textit{E-mail: obridko@izmiran.ru}}
\centerline{\textit{$^2$IZMIRAN, Troitsk, Moscow, Russia}}
\centerline{\textit{$^3$Sternberg Astronomical Institute of the Lomonosov Moscow State University, Moscow, Russia}}
\smallskip
\centerline{\small Received July 8, 2023}
\smallskip
\textbf{Abstract.} 
Further development of the work of Obridko et al. [1] based on recent data 
confirms the assumption that the 25th cycle of solar activity is a 
medium-low cycle. Its height is expected to be $125.2\pm5.6$, and the expected 
date of the maximum phase is the end of 2023 or the first quarter of 2024.

\section*{Introduction}

Obridko et al. [1] analyzed the evolution of the large-scale magnetic field 
on the Sun during the last four cycles from 1976 to early 2022.  WSO data
(\url{http://wso.stanford.edu/}) have been used. Special attention 
was paid to the effect of prolonged cycles of solar activity. The term appeared 
in literature in 1988 [2, 3], although during that decade, observational 
evidence appeared indicating that the magnetic activity of one cycle 
overlapped for some period of time (often up to several years) with that of 
the previous cycle [4]. As observed on the surface, the extended solar 
cycle starts during the sunspot maximum at high latitudes and consists of a 
relatively short polarward branch (described as "rush to the poles") and a 
long equatorward branch, which continues through the solar minimum and the 
following sunspot cycle [5], see also [6] for review.

The polaward and equatorward waves appear almost simultaneously and have 
opposite predominant polarity of the magnetic field. There are periods when 
two “rush to the poles” waves with the field of opposite signs coexist in 
the Sun, one of which has nearly reached the pole and the other has just 
appeared at mid latitudes. In such periods we can see three zones of 
alternating polarity in each hemisphere.

The moment, when all three types of waves are simultaneously present on the 
disk and, as a whole, six intermittent zones are observed, exactly 
corresponds to the zonal harmonic with $\it l=5$. We propose to call this 
time interval "the overlapping phase". As seen below, the overlapping phase 
can be quantitatively described in terms of the 5th zonal magnetic field 
harmonics, which, in this connection, can be referred to as the height of 
the overlapping phase. During the overlapping phase, three activity waves 
coexist on the solar surface, which lead to the appearance and enhancement 
of the odd zonal harmonic with $\it l = 5$.

\begin{figure}[h!]
\begin{center}
\includegraphics*[height=.5\textheight]{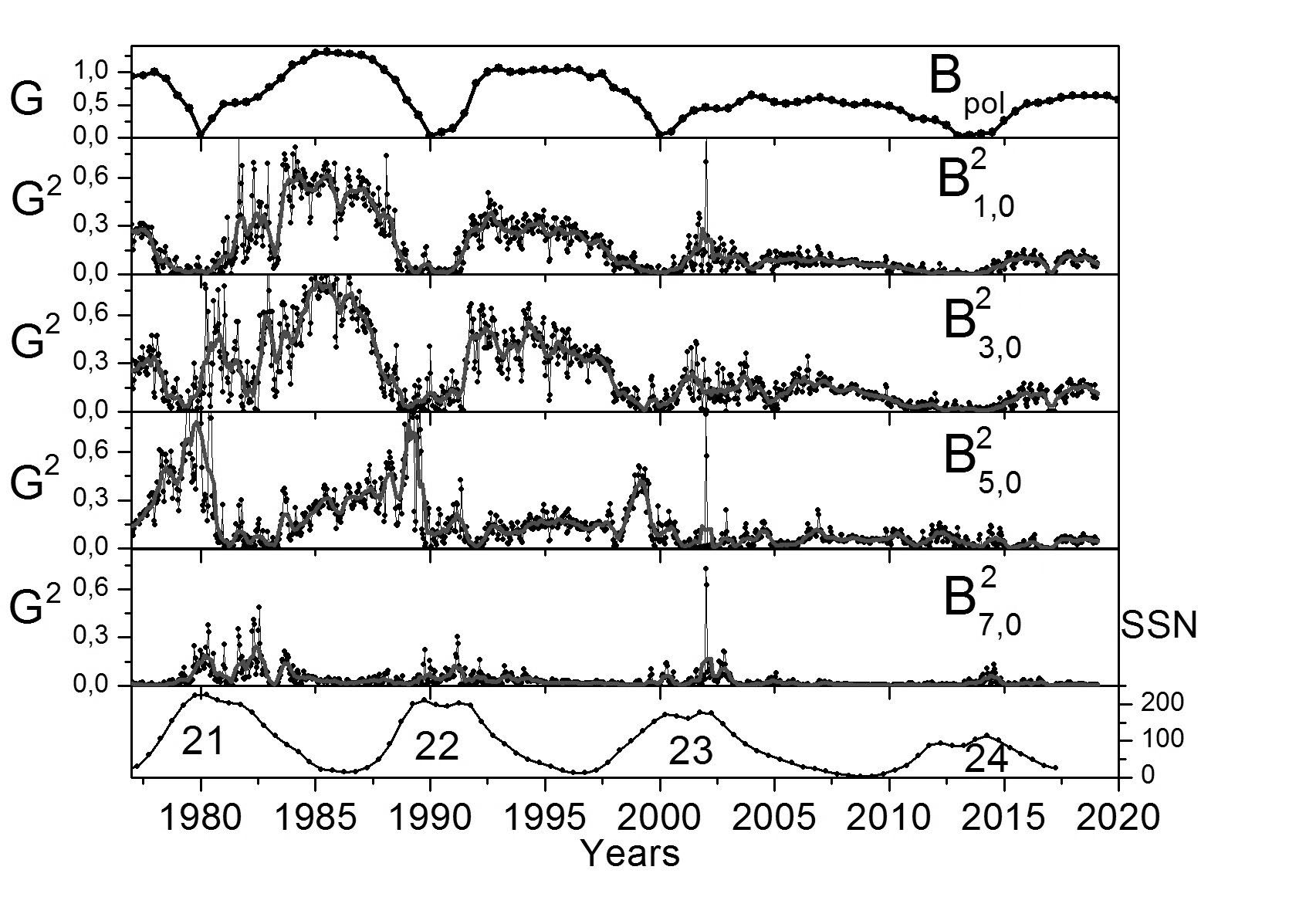}
\end{center}
\caption{Time dependence of the squared magnetic field connected with the first odd axisymmetric harmonics up to $\it l = 7$ [1]
. The lowest plot represents the time variation of sunspot numbers (SSN) with the cycle numbers indicated; the top plot is the 
polar field.}
\end{figure}

The maximum amplitude of this harmonic dramatically decayed over the past 
four cycles similar to the cycle amplitude recorded in sunspot numbers. A 
particularly strong decline in the value of the $g_{5,0}$ harmonic is 
observed after 2000, and this led to the low Cycle 24. Of course, four 
cycles are not enough to provide a convincing statistics; however, it seems 
plausible that Cycle 25 will not be much higher than Cycle 24.

\section*{New estimates}

At present, new data have appeared that support this conclusion.

\begin{enumerate}
\item In the middle of 2022, an increase of the fifth zonal harmonic was recorded. Although this peak is not high and is much lower than in Cycles 21, 22, and 23, it certainly indicates the near onset of a low sunspot maximum. In Cycles 21, 22 and 23, the maximum of this harmonic was ahead of the sunspot maximum by no more than 1-1.5 years.
\item In March 2023, a reversal of the polar magnetic field was recorded, and in June 2023, half the sum of the field values at both poles turned to zero. This usually indicates the proximity of the sunspot maximum. At the same time, in Cycles 21 and 22, the field reversal virtually coincided with the date of the sunspot maximum; in the low Cycle 24, the polarity reversal was ahead of the sunspot maximum by about a year.
\item The relation between the magnitude of the polar field and the height of the upcoming sunspot maximum number has been often used for forecasting. At present, reliable measurements of the polar field strength in Cycles 22, 23, and 24 are available. The relationship between the polar field strength and the height of the upcoming cycle is described by the formula:
\end{enumerate}

\begin{equation}
SSN_{max}=36.405+1.3666 \, B_{pol}
\end{equation}

\begin{figure}[h!]
\begin{center}
\includegraphics*[height=.5\textheight]{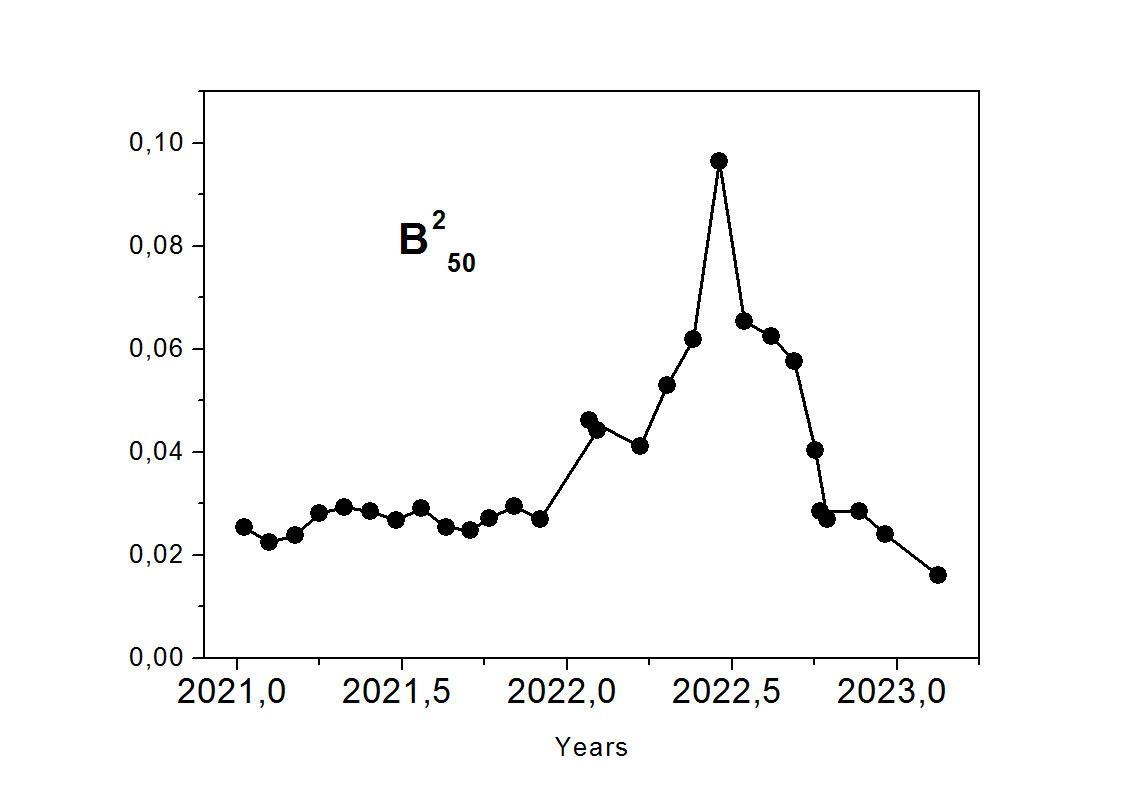}
\end{center}
\caption{The squared amplitude of the fifth harmonic as a function of time 
in 2021-2022.}
\end{figure}

The last maximum value of $B_{pol}$ equal to 65 $\mu T$ was recorded in 
summer 2019 (see the site \url{http://wso.stanford.edu/Polar.html}). Hence, 
the predicted value of SSN in Cycle 25 is $125.2\pm5.6$. This is only a few 
units higher than in Cycle 24 (116.4).

\medskip
The authors are grateful to Dr. T.Hoeksema for access to the the site 
\url{http://wso.stanford.edu}
We acknowledge the support of the Ministry of Science and Higher Education 
of the Russian Federation under the grant 075-15-2020-780 (VNO and MMK) and 
075-15-2022-284 (DDS).

\section*{References}

\hspace{6.5mm}1.Obridko V. N., Shibalova A. S., and Sokoloff, D. D.,  MNRAS \textbf{523},  1,  982 (2023).

2. Altrock R. C., in: Solar and Stellar Coronal Structure and Dynamics, ed. by R.C.Altrock. pp 414--420 (1988).

3. Wilson P. R., Altrock R. C., Harvey K. L., Martin S. F., and Snodgrass H. B., Nature \textbf{333}, 748 (1988).

4. Leroy J. L. and Noens J. C., A\&A \textbf{120}, L1 (1983).

5. Kosovichev A., Pipin V., and Getling A.,  in: American Astronomical Society Meeting Abstracts. p. 304.05 (2021).

6. McIntosh S. W. et al.,  Sol. Phys. \textbf{296}, 189 (2021).

\end{document}